\documentclass[aps,prl,reprint,showpacs,twocolumn,superscriptaddress]{revtex4}

\usepackage[dvips]{graphicx}
\usepackage{dcolumn}
\usepackage{bm}
\usepackage{xspace}
\usepackage{multirow}
\usepackage{natbib}
\usepackage{color}

\begin{document}

\preprint{}
\title{Measurement of an Exceptionally Weak Electron-Phonon Coupling on the
   Surface of the Topological Insulator Bi$_2$Se$_3$ Using Angle-Resolved
   Photoemission Spectroscopy}

\author{Z.-H. Pan}
\affiliation{Condensed Matter Physics and Materials Science Department, Brookhaven National Lab, Upton, NY 11973}
\author{A. V. Fedorov}
\affiliation{Advanced Light Source, Lawrence Berkeley National Laboratory, Berkeley, CA 94720}
\author{D. Gardner}
\author{Y.S. Lee}
\affiliation{Department of Physics, Massachusetts Institute of Technology, Cambridge, MA 02139}
\author{S. Chu}
\affiliation{Center for Materials Science and Engineering, Massachusetts Institute of Technology, Cambridge, MA 02139}
\author{T. Valla}
\email{valla@bnl.gov}
\affiliation{Condensed Matter Physics and Materials Science Department, Brookhaven National Lab, Upton, NY 11973}
\date{\today}

\begin{abstract}
Gapless surface states on topological insulators are protected from elastic scattering on non-magnetic impurities which makes them promising candidates for low-power electronic applications.  However, for wide-spread applications, these states should have to remain coherent at ambient temperatures. Here, we studied temperature dependence of the electronic structure and the scattering rates on the surface of a model topological insulator,  Bi$_2$Se$_3$, by high resolution angle-resolved photoemission spectroscopy. We found an extremely weak broadening of the topological surface state with temperature and no anomalies in the state's dispersion, indicating exceptionally weak electron-phonon coupling. Our results demonstrate that the topological surface state is protected not only from elastic scattering on impurities, but also from scattering on low-energy phonons, suggesting that topological insulators could serve as a basis for room temperature electronic devices.

\end{abstract}
\vspace{1.0cm}


\maketitle
\pagebreak
Three-dimensional topological insulators (TIs) have Dirac-like surface states in 
which the spin of the electron is locked perpendicular to its momentum
 in a chiral spin-structure where electrons with opposite momenta
have opposite spins \cite{Fu2007a,Noh2008,Hsieh2008,Zhang2009,Hsieh2009,Xia2009,Chen2009,Pan2011}.
A direct consequence of the chiral spin-structure is that a 
backscattering, which would require a spin-flip process, is not allowed if a time-
reversal-invariant perturbation, such as non-magnetic disorder, is present 
\cite{Fu2007a}, making these surface states promising candidates for spintronics and quantum 
computing applications, where the spin-coherence is crucial.  \cite{Biswas2010,Fu2009,Guo2010,Liu2009,Zhou2009,Chen2010a,Wray2010}. 
Recent scanning tunneling microscopy (STM) experiments 
\cite{Roushan2009,ZhangSTM2009,Alpichshev2010,Seo2010,Hanaguri2010} have 
shown that backscattering is indeed strongly suppressed or completely 
absent, despite strong atomic scale disorder. 
Our own angle-resolved photoemission spectroscopy (ARPES) studies have indicated that the state is remarkably insensitive to both non-magnetic and magnetic impurities in the low doping regime, where the 
Fermi surface (FS) is nearly circular. The scattering is found to increases as the FS becomes 
hexagonally warped with increased doping, irrespective of the impurity's magnetic moment \cite{Valla2012}. 

While the elastic scattering imposes the ultimate limit on the charge transport,  the inelastic scattering processes dictate material's transport properties at finite temperatures. In particular, interactions of electrons with  lattice modes is responsible for increasing resistivity with temperature in metals. The same interaction may also lead to a ground states with broken 
symmetries, such as superconductivity or charge-density-wave state. So far, inelastic scattering processes at surfaces of TIs have been scarcely studied, with only one theoretical study on the coupling of the topological surface states  (TSS) to phonons \cite{Giraud2011}. As the lattice modes in general do not represent a time-reversal symmetry breaking perturbation, it might be expected that the TSS should not couple to the $q\approx 2k_F$ phonons. Therefore, scattering on phonons should resemble scattering on non-magnetic impurities, where the rates are shown to be sensitive to the Fermi surface size and shape \cite{Valla2012}. On the other hand, the proximity of bulk states, which in some cases could be strongly coupled to phonons (occurrence of superconductivity upon Cu-doping in Bi$_2$Se$_3$ and under pressure in Bi$_2$Te$_3$ \cite{Hor2010,Zhang2011,Kriener2011}), could also influence the TSS by allowing the inter-band electron-phonon scattering. As these processes will play a crucial role in determining performances of any real devices based on TIs, their better understanding is an imperative. 

In this Letter, we present the high resolution ARPES studies of the scattering rates on the surface of a  TI,  Bi$_2$Se$_3$. We observe a very weak temperature broadening of the TSS and no anomaly in the state's dispersion due its coupling to phonons. Our results show that the electron-phonon coupling is suppressed  in a similar way as the elastic scattering, suggesting that TIs could serve as a basis for room temperature applications.

The experiments were carried out on a Scienta SES-100 electron spectrometer 
at the beamline 12.0.1 of the Advanced Light Source (ALS) and on a Scienta 2002 analyzer at the beamline U13UB of the National Synchrotron Light Source (NSLS). The spectra were recorded at the photon energy of 50 eV and 18.7 eV,
with the combined instrumental energy resolution of $\sim12$ meV and $\sim8$ meV, at ALS and NSLS, respectively. The angular resolution was better than $\pm 0.07^{\circ}$ in both instruments. The single crystals of Bi$_2$Se$_3$ were synthesized by mixing stoichiometric amounts of bismuth and selenium with trace amounts of arsenic in evacuated quartz tubes \cite{Steinberg2010}. Samples were cleaved at low temperature (15-20 K) under ultra-high vacuum (UHV) conditions ($2\times10^{-9}$ Pa). The temperature was measured using a silicon sensor mounted near the sample.

\begin{figure}[htb]
\begin{center}
\includegraphics[width=8cm]{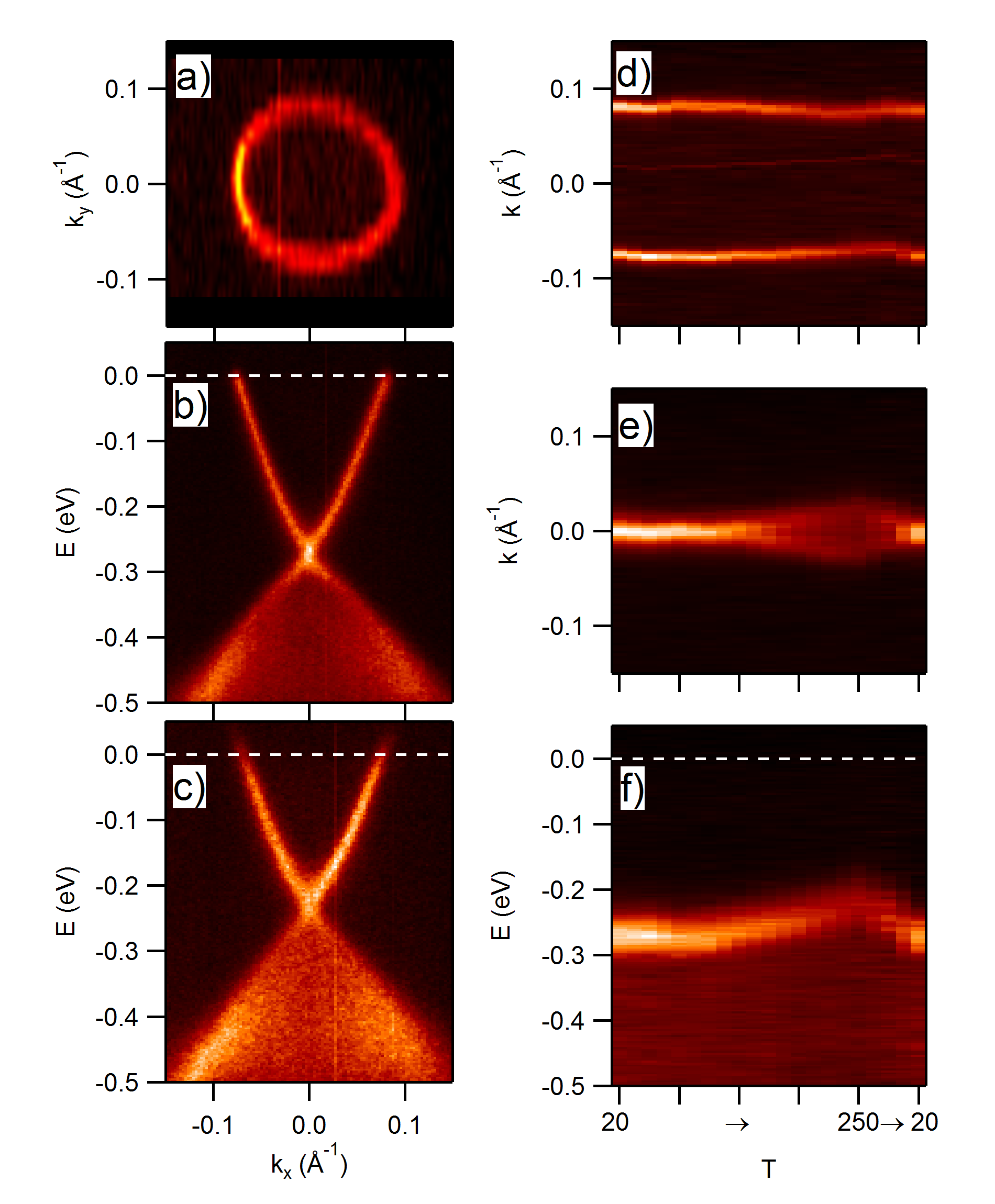}
\caption{Temperature effects on the ARPES spectra from Bi$_2$Se$_3$. (a) Fermi surface of  Bi$_2$Se$_3$ at 18 K. (b) ARPES intensity along the $\Gamma$K line in the surface Brillouin zone at 18K and (c) at 255 K. Photoemission intensity at the Fermi level (d) and at $E=-270$ meV (e) along the $\Gamma$K momentum line as a function of temperature. (f) Photoemission intensity at the $\Gamma$ point as a function of temperature. Sample was heated from 18 K to 255 K and then cooled back to 18 K.
}
\label{Fig1}
\end{center}
\end{figure}
Fig. \ref{Fig1} illustrates the effects of raising temperature on the electronic structure of Bi$_2$Se$_3$ measured in ARPES around the center of the surface Brillouin zone. The rapidly dispersing conical state in Fig. 1b) and c) represents the TSS that forms a circular Fermi surface shown in Fig. 1a. Its filling varies with temperature as evident from the shift of the Dirac point from  $\approx0.27$ eV below the Fermi level at 18 K to $\approx0.23$ eV at 255 K. This temperature induced shift and the corresponding change in the Fermi surface area are fully reversible upon temperature cycling as can be seen in panels d) to f).  We note that at the pressure of $2\times10^{-9}$ Pa, TSS is very stable if kept at constant temperature, without noticeable changes in the spectra several hours after cleaving. Therefore, the effects shown in Fig. 1 reflect the intrinsic temperature induced changes in the quasi-particle dynamics rather than some spurious effects caused by adsorption/desorption of residual gases. We note that similar shifts in binding energy of the state with temperature were observed in Shockley-type surface states on noble metals. These shifts could be explained in the simple phase accumulation model where the phase change on the crystal side of the potential well, that determines the energy of the surface state, is affected by slight changes in the bulk band gap as temperature is varied \cite{Paniago1995}. In the case of Bi$_2$Se$_3$, the  bulk valence band (BVB) is expected to have the dominating effect on the energy of the Dirac point. The upward shift of the Dirac point, would indicate that the BVB also shifts up and that the bulk band gap in Bi$_2$Se$_3$ decreases with increasing temperature.

\begin{figure*}[htb]
\begin{center}
\includegraphics[width=13cm]{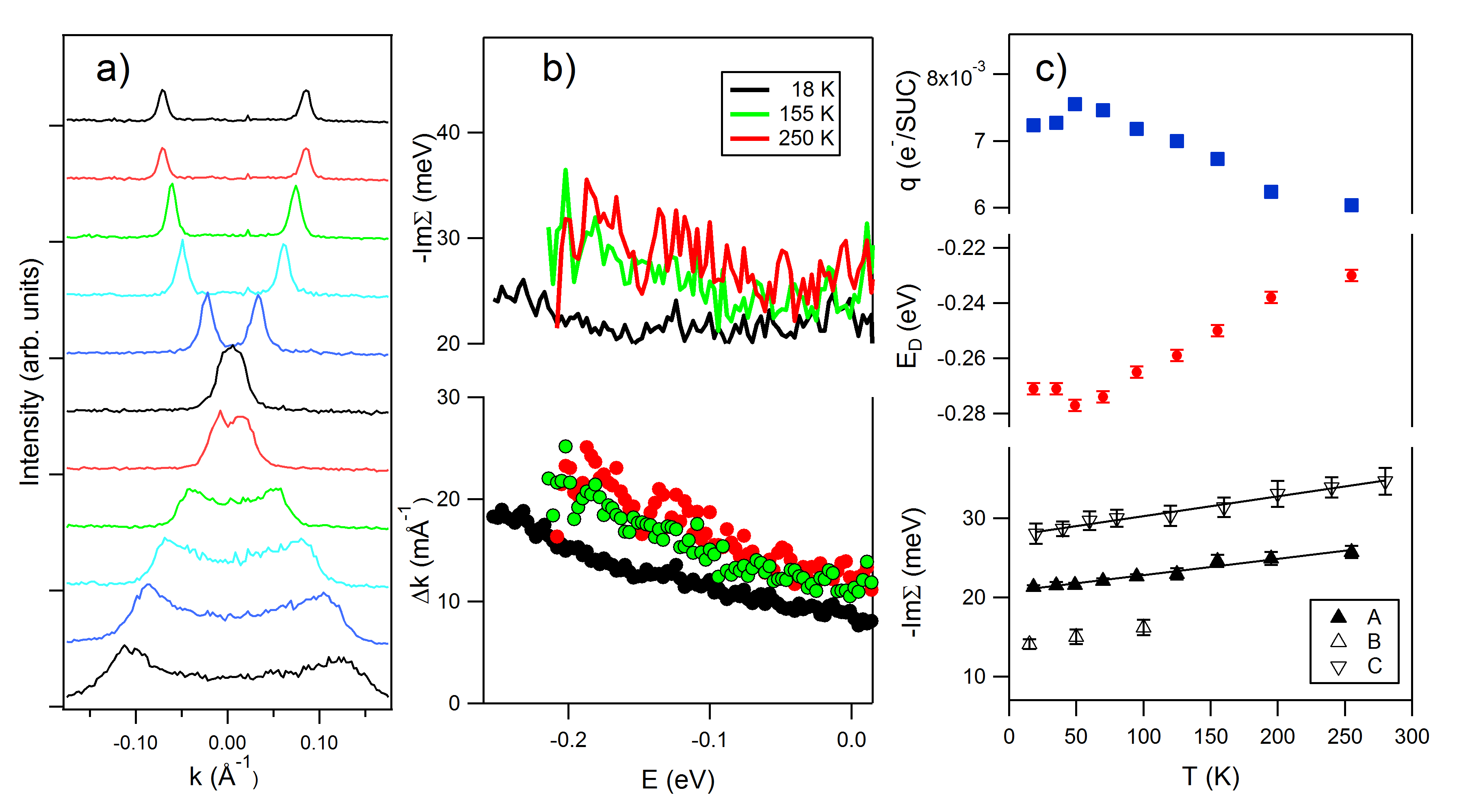}
\caption{Temperature broadening of TSS on Bi$_2$Se$_3$. (a) Momentum distribution curves (MDCs) corresponding to the spectrum shown in Fig. 1(b), spaced by 50 meV, with the top curve representing the Fermi level. (b) Momentum width $\Delta k$ (bottom) and Im$\Sigma$ (top) of the Lorentzian-shaped MDC peaks at several different temperatures. Standard deviations from the fitting are within 5\% of the obtained value (not shown)  (c) Temperature dependence of Im$\Sigma(0)$ for three different samples (bottom), $E_D$ (middle) and doping level of TSS for sample A (top). Solid lines are the linear fits of Im$\Sigma(0)$.
}
\label{Fig2}
\end{center}
\end{figure*}

To quantify the changes in the spectral width of TSS, we have analyzed the photoemission spectra at different temperatures using the standard method where the momentum distribution curves (MDCs) are fitted with Lorentzian peaks \cite{Valla1999,Valla1999b}. The width of the Lorentzian peak, $\Delta k(\omega)$, is related to the quasiparticle scattering rate $\Gamma(\omega)=2|$Im$\Sigma(\omega)|=\Delta k(\omega)v_0(\omega)$, where $v_0(\omega)$ is the bare group velocity and Im$\Sigma(\omega)$ is the imaginary part of the complex self-energy. Fig. 2 shows several MDCs corresponding to the spectrum from Fig. 1b) and summarizes the results of the analysis. The spectral region above the Dirac point is very clean: it consists of two Lorentzian-shaped peaks with essentially no background intensity, the fact that makes the fitting procedure very accurate. The bulk conduction band is absent as at the chosen photon energy of 50 eV, that corresponds approximately to the $Z$ point in the bulk Brillouin zone, it lays above the Fermi level. In contrast, the spectral region below the Dirac point is always affected by the BVB. The fitting results for the region above the Dirac point are shown in panel b). Im$\Sigma$ displays a weaker energy dependence than $\Delta k$, reflecting an increasing group velocity as the state approaches the Fermi level. However, the most important observation here is that Im$\Sigma$ near the Fermi level shows very little change between 18 K and 255 K. Temperature broadening of a quasi-particle peak usually reflects an increase in the scattering on phonons and its near absence here points to a very weak coupling of TSS to phonons in Bi$_2$Se$_3$.
The electron-phonon coupling constant, $\lambda$, can be determined from the temperature slope of Im$\Sigma(0)$ because at higher temperatures, approximately $k_BT>\Omega_0/3$, the electron-phonon self energy
\begin{equation}
|{\mathrm Im}\Sigma(\omega,T)|=\pi \int_{0}^{\infty}d\nu \alpha^2F(\nu)[2n(\nu)+f(\nu+\omega)+f(\nu -\omega)]
\label{eq1}
\end{equation}    
is approximately linear in temperature, Im$\Sigma(0,T)\approx \lambda\pi k_BT$. Here $\alpha^{2}F(\omega)$ is the Eliashberg coupling function, $f(\omega)$ and $n(\omega)$ are the 
Fermi and Bose-Einstein functions, $\Omega_0$ is energy of the highest involved phonon and $k_B$ is Boltzmann's constant \cite{Grimvall}. In panel c), we plot Im$\Sigma$ averaged over -20 meV $<\omega<0$ as a function of temperature for three different samples. The error bars represent the standard deviation of the averaged value. Samples A and C were measured at 50 eV, while sample B was measured at 18.7 eV photon energy. Differences in the TSS's width are partially due to the different momentum resolution at these two photon energies and partially due to the natural variation in the surface "quality". However, in all three samples Im$\Sigma$ increases with temperature at a similar rate. The increase starts at low temperatures, indicating the involvement of low energy phonons. The linear fits give $\lambda=0.076\pm 0.007$ for sample A and $0.088\pm 0.009$ for sample B. This represents one of the weakest coupling constants ever reported in any material, weaker than the theoretical value from ref. \cite{Giraud2011}, but in agreement with the apparent absence of temperature broadening of TSS in recent experiments on several topological materials \cite{Park2011}. In contrast, the occurrence of superconductivity upon Cu-doping in Bi$_2$Se$_3$ and under pressure in Bi$_2$Te$_3$ \cite{Hor2010,Zhang2011,Kriener2011} suggests much stronger coupling in the bulk of these materials. The estimate for the bulk coupling constant can be made by using the known values for Debye temperatures \cite{Shoemake1969} and superconducting transition temperatures ($T_c$) \cite{Hor2010,Zhang2011,Kriener2011} in McMillan's formula for $T_c$ \cite{McMillan1968}: $\lambda=0.62$ (0.6) is obtained for Cu$_x$Bi$_2$Se$_3$ (Bi$_2$Te$_3$), almost an order of magnitude stronger than our result for the surface state. 

\begin{figure}[hb]
\begin{center}
\includegraphics[width=6.5cm]{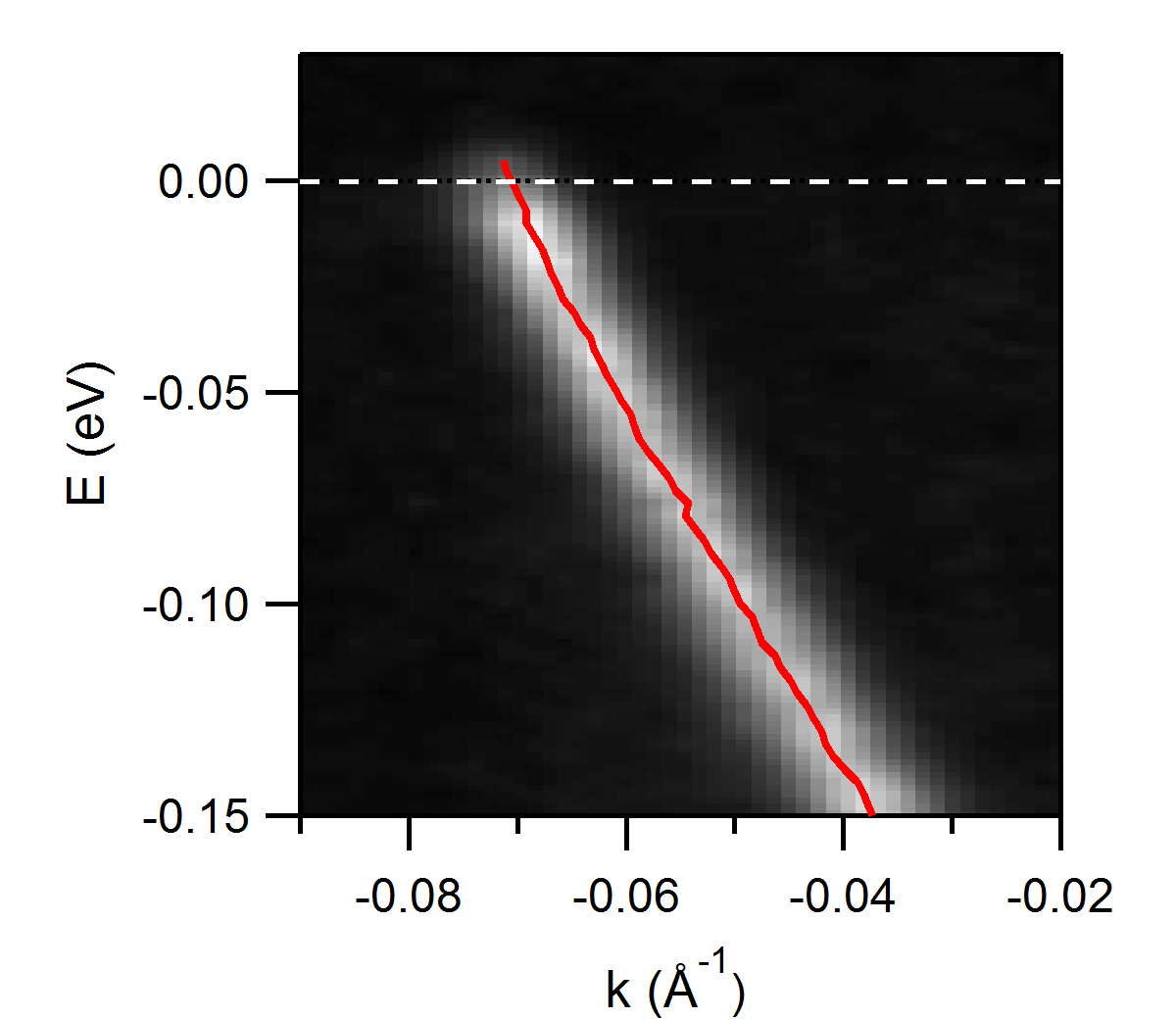}
\caption{Zoom in the low-energy region of the ARPES spectrum from Bi$_2$Se$_3$ from Fig. 1b). Dispersion of TSS (solid line) is obtained from positions of Lorentzian-fitted peaks in MDCs. 
}
\label{Fig3}
\end{center}
\end{figure}

Another indication of the exceptionally weak electron-phonon coupling at the surface is the apparent absence of a mass enhancement in the dispersion of TSS near the Fermi level. In Fig. 3 we show the low energy region of the ARPES spectrum from Fig. 1b).  A hallmark of the quasiparticle coupling to phonons in the form of a sudden change in the slope or a "kink" in dispersion inside the phonon-energy range \cite{Valla1999b} is conspicuously missing. The MDC derived dispersion in Fig. 3 is essentially a straight line with no anomalies in the vicinity of the Fermi level, in agreement with previous studies \cite{Park2011}.  We note that the temperature dependence from Fig. 2c) requires the involvement of low energy modes which, in addition to the very weak coupling, makes the observation of an anomaly in dispersion extremely difficult and it would require a much better experimental resolution. We also note that the finite experimental resolution probably already affects the extracted values of Im$\Sigma$ at low temperatures and that $\lambda$ obtained from temperature dependence might be slightly underestimated. 

Our results should have very important consequences on the macroscopic properties of the Bi$_2$Se$_3$ surface, in particular on the surface state's contribution to transport - a crucial aspect for any (spin) electronic device based on TSS. The surface contribution to transport has proven elusive due to the overwhelming bulk component to conductivity and/or low surface state mobility in the 
environment of a typical transport measurement \cite{Butch2010,Analytis2010a,Analytis2010b,Qu2010,Eto2010}. 
\begin{figure}[htb]
\begin{center}
\includegraphics[width=6.5cm]{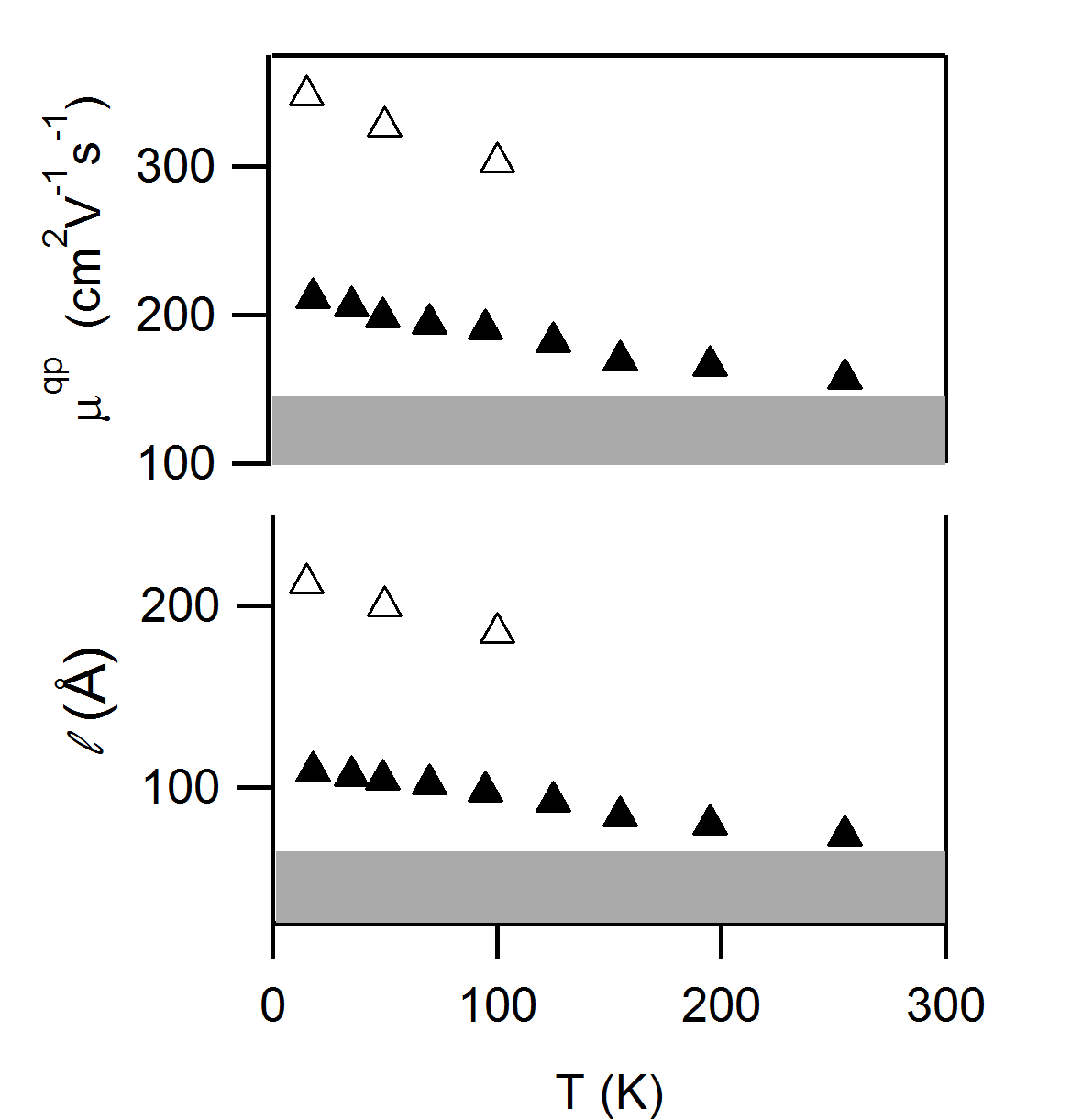}
\caption{Quasiparticle mean-free path $\ell$ (bottom) and $\mu^{qp}$ (top) as functions of temperature, determined from the ARPES spectra for samples A and B from Fig. 2c). Gray regions represent the limits of these quantities in doped surfaces \cite{Valla2012}. 
}
\label{Fig4}
\end{center}
\end{figure}
The determining factor for transport is the surface state mobility, which can be expressed as $\mu_S=e\ell_{tr}/(\hbar k_F)$ for the Dirac-like carriers. Here, $\ell_{tr}$, represents the transport mean free path. In ARPES experiments, $k_F$ and the quasiparticle mean-free path $\ell=(\Delta k)^{-1}$ can be directly measured. In Fig. 4, we plot the quasiparticle mean free path and the quantity $\mu^{qp}=e\ell/(\hbar k_F)$, which may serve as a lower bound for surface state mobility, as functions of temperature for samples A and B. We note that $\ell_{tr}$ might be significantly longer than $\ell$, because currents in general are not sensitive to the small angle scattering events that may dominate $\ell$. This discrepancy might be especially enhanced in systems in which the backscattering is suppressed, as in the case of TSSs, and we might expect significantly higher mobilities than $\mu^{qp}$ shown in Fig. 4. Therefore, the unperturbed and strongly coherent TSSs, as those measured here, have a strong potential to serve as a basis for room temperature spintronic devices. However, the environmental exposure will inevitably affect the coherence of the topological state and degrade its mobility, in a similar way as it was demonstrated in ref. \cite{Valla2012}. We note that recent transport experiments have detected quantum oscillations related to the TSS, yielding surface mobilities of around $10^4$ cm$^2$V$^{-1}$s$^{-1}$ on the surface of Bi$_2$Te$_3$ \cite{Qu2010}, still low compared to those measured in suspended graphene or in the best semiconductors \cite{Harris1987,Bolotin2008}. We suggest that controlled (ultra-high vacuum) environment and/or an inert capping of the surface would further improve the mobilities of TSS and that such measures might be necessary for optimal functioning of TI-based devices.

In summary, we have observed a weak electron-phonon coupling on the surface of Bi$_2$Se$_3$ demonstrating that TSS is well protected from scattering on low-energy phonons. This keeps the possibility that TSSs could serve as a basis for room-temperature devices open.

The work at Brookhaven is supported by the US Department of Energy (DOE) under Contract No. DE-AC02-
98CH10886. The work at MIT is supported by the DOE under Grant No. DE-FG02-04ER46134.
ALS is operated by the US DOE under Contract No. DE-AC03-76SF00098.

\end{document}